\documentclass[aps,prl,reprint,superscriptaddress,longbibliography]{revtex4-2}
\usepackage{multirow}
\usepackage{graphicx}
\usepackage{enumerate}
\usepackage{enumitem}
\usepackage{amsmath}
\usepackage{amssymb}
\usepackage{stix}
\usepackage{mathtools}
\usepackage{wasysym}
\usepackage{fontawesome}

\usepackage{float}

\begin{document}

\def\crta{\vrule height1.41ex depth-1.27ex width0.34em}
\def\dj{d\kern-0.36em\crta}
\def\Crta{\vrule height1ex depth-0.86ex width0.4em}
\def\Dj{D\kern-0.73em\Crta\kern0.33em}
\dimen0=\hsize \dimen1=\hsize \advance\dimen1 by 40pt

\title{On Simplest Kochen-Specker Sets}

\author{Mladen Pavi\v ci\'c}
\email{mpavicic@irb.hr}

\affiliation{Center of Excellence for Advanced Materials 
and Sensing Devices (CEMS), Photonics and Quantum Optics Unit,\\ 
Ru{\dj}er Bo\v skovi\'c Institute, Zagreb, Croatia.}

\begin{abstract}
  In {\em Phys. Rev. Lett.\/} {\bf 135}, 190203 (2025) a discovery of
  the simplest 3D contextual set with 33 vertices, 50 bases, and 14
  complete bases is claimed. In this paper, we show that it was
  previously generated in {\em Quantum\/} {\bf 7}, 953 (2023) and
  analyze the meaning, origin, and significance of the simplest
  contextual sets in any dimension. In particular, we prove that there
  is no ground to consider the aforementioned set as fundamental
  since there are many 3D contextual sets with a smaller number of
  complete bases. We also show that automatic generation of contextual
  sets from basic vector components automatically yields all known
  minimal contextual sets of any kind in any dimension and therefore
  also the aforementioned set in no CPU-time. In the end, we discuss
  varieties of contextual sets, in particular Kochen-Specker (KS),
  extended KS, and non-KS sets as well as ambiguities in their
  definitions.
\end{abstract}

 \maketitle

 Recently a number of experiments \cite{pavicic-quantum-23} paved
the road of possible applications of contextual sets in quantum
computation \cite{magic-14,bartlett-nature-14}, quantum steering
\cite{tavakoli-20}, and quantum communication \cite{saha-hor-19}.
Under a contextual set we understand a quantum set to whose
elements an assignment of predetermined (classical) 0--1 values
is impossible but which nevertheless allow consistent 0--1 outcomes
within a quantum measurement.

Such contextual sets might be represented by graphs, hypergraphs,
operators, projectors, states, vectors, matrices, etc. Our focus
is on special kind of general hypergraphs
\cite{berge-73,berge-89,bretto-13,voloshin-09} which are called
McKay-Megill-Pavi\v ci\'c hypergraphs (MMPH)
\cite{pavicic-quantum-23}.

A hypergraph is a set of points and a set of subsets of these
points. The points are called the vertices of the hypergraph
and the subsets are called the hyperedges of the hypergraph.
Vertices might be represented by vectors, operators, subsets, or
other objects, and hyperedges by a relation between vertices
contained in them such as orthogonality, inclusion, or geometry.
MMPH is defined in \cite[Def.~2.1]{pavicic-quantum-23}.
Contextuality of MMPHs is defined as follows.

{\em Def.}  A $k$-$l$ {\rm MMPH} of dim $n$$\ge$$3$ ($n$ is the
max No.~of vertices in hyperedges) with $k$ vertices and $l$
hyperedges, whose $i$-th hyperedge contains $\kappa(i)$ vertices
2$\le$$\kappa(i)$$\le$$n$, $i$=1,\dots,$l$ to which it is impossible
to assign {\rm 1}s and {\rm 0}s in such a way that (i) no two
vertices within any of its hyperedges are both assigned the value
$1$ and (ii) in any of its hyperedges, not all of the vertices are
assigned the value $0$, is a {\em contextual\/} {\rm MMPH}.

{\em Lemma.} A contextual MMPH whose vertices are represented by
vectors and hyperedges defined by their orthogonalities is a {\em
Kochen-Specker\/} (KS) contextual set provided each of its
hyperedges contains $n$ vertices and a non-KS contextual set
provided at least one of its hyperedges contains less than $n$
and at least two vertices.~\cite[Theorem 3.1]{pavicic-quantum-23}

{\em Def.\/} We say that vertices which belong to $m$ hyperedges have
the vertex {\em multiplicity\/} $m$. 

{\em Def.\/} A contextual MMPH whose removal of any of its hyperedges
turns it into a non-contextual MMPH is called a {\em critical\/} MMPH. 

{\em Def.\/} A {\em master\/} MMPH is a non-critical MMPH
that contains smaller critical and non-critical sub-MMPHs.
A collection of all sub-MMPHs of an MMPH master forms its
{\em class}.

\begin{figure}[h]
\begin{center}
  \includegraphics[width=0.49\textwidth]{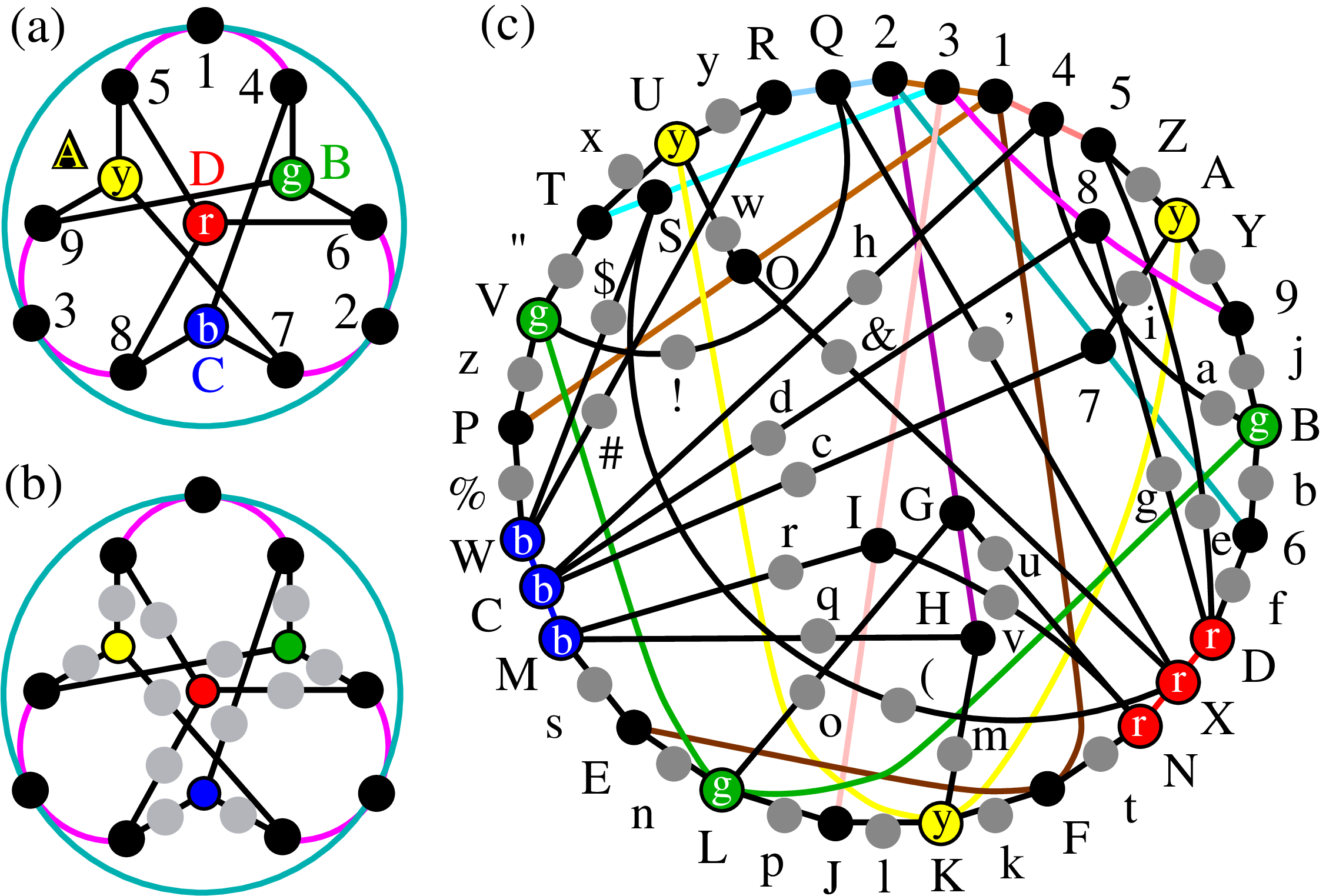}
\end{center}
\caption{(a) The Yu-Oh set 13-16 \cite{yu-oh-12} in an MMPH
  representation; (b) the Yu-Oh set filled with grey vertices
  of multiplicity 1---the 25-16 MMPH; (c) the 69-50 MMPH from
  \cite[Fig.~10(e)]{pavicic-quantum-23} redrawn so as to flash
  colored vertices from \cite[Fig.~1]{cabello-25b}; its variety
  with grey vertices of multiplicity 1 dropped---33-50---is
  isomorphic to \cite[Fig.~1]{cabello-25b} and to 
  Fig.~\ref{fig:yuoh} below; its coordinatization is given in
  the Appendix.} 
\label{fig:CK}
\end{figure}

Some authors call non-KS sets KS sets and KS sets extended KS
sets \cite{larsson,cabello-25b}. They  hold that ``every extended
KS set is an original KS set'' \cite[p.~1]{cabello-25b}. The
statement does not hold for, e.g., the contextual set 13-16 
shown in Fig.~\ref{fig:CK}(a) vs.~its extended non-contextual 25-16
set shown in Fig.~\ref{fig:CK}(b). Also there is a terminological
ambiguity in generally accepted notation in higher dimensions, e.g.,
for the 4D 18-9 set as we show below. 

Here we have some problems, though. First, the historically known
minimal 3D sets are not critical contextual sets, while their
``extended'' sets are. \cite[p.~8, Fig.~4]{pavicic-entropy-19}.
We guess that the authors (Kochen, Specker, Bub, Conway) were
well aware that together with any two original vectors in 3D
there is a third vector orthogonal to both of them
(notwithstanding whether one takes it into account or not within
a calculation), but that they dropped such vertices of
multiplicity 1 just to make their sets appear smaller. 
Bub explicitly stated that, since he started with an ``extended''
KS 49-36 set to finally arrive at the ``simplest'' 33-36 KS
set: ``Removing \dots\ 16 rays \dots  that occur in only one
orthogonal triple \dots\ from the 49 rays yields \dots [a] set
of 33 rays.'' \cite{bub}

Second, an effort of minimising sets and reaching records
is dispensable since they all come out automatically from
basic vector components. In
Refs.~\cite{pm-entropy18,pavicic-pra-22,pw-23a} we show that
basic vector components generate classes of KS (extended KS)
MMPHs (sets) in any dimension up to 32 and higher and that
the classes contain any known KS set and any known minimal KS
instance from the literature. So, in 3D, vector components
$\{0,\pm 1,\pm 2, 5\}$ yield the 97-64 class with 20 critical
MMPHs which include ``extended'' Bub's set, 9 non-isomorphic
51-37 MMPHs one of which is Conway-Kochen's set, a 53-38, 8
54-39, and a 55-40 \cite[Table I]{pavicic-pra-22},
$\{0,\pm 1,\pm\sqrt{2},3\}$ yield the 81-52 class which contains
a single critical set---``extended'' Peres' set, a set of 24
vector components explicitly given in
\cite[Supp.~Material, p.3]{pavicic-pra-22} yields ``extended''
Kochen-Specker's set, and
$\{0,\pm\omega,2\omega,\pm\omega^2,2\omega^2\}$, where
$\omega=e^{2\pi i/3}=(-1+i\sqrt{3})/2$, yield the 169-120 class
which contains the minimal critical set 69-50 explicitly given
in \cite[Fig.~10(e), p.~54]{pavicic-quantum-23}. In
Fig.~\ref{fig:CK}(c) it is redrawn for a better transparency. 

Now, when we drop the vertices with multiplicity 1 (grey dots)
from the 69-50 set \cite[Fig.~10(e), p.~54]{pavicic-quantum-23}
we obtain a 33-50 set (cf.~the aforementioned Bub's procedure),
which is isomorphic to the ``new record'' set
\cite[Fig.~1]{cabello-25b} Cabello obtained two years later.

Hence, the set of \cite{cabello-25b} was known previously.

Let us now consider some other points.

{\em 3D presentation.\/} \ Fig.~1 from \cite{cabello-25b} offers
a narrative description on how one can redraw the 33-50 set
(65-50 with grey vertices in Fig.~\ref{fig:CK}(c) dropped) in a
real three-dimensional space. However, this is inconsistent since
we deal with complex vectors and therefore if we wanted to put the
set in a real space it should be a six-dimensional space. 
With an MMPH representation of the set we do not have this problem
because its dimensionality is defined by the maximal number of
vertices within its hyperedges. We realize the representation
be means of a model implemented by Pavi{\v c}i{\'c} Ravli{\'c}
via Blender 3D graphics suite which enables the reader to
interactively view the model from a chosen
angle.~\cite{pavicic-ravlic-2025}

\begin{figure}[ht]
\begin{center}
  \includegraphics[width=0.49\textwidth]{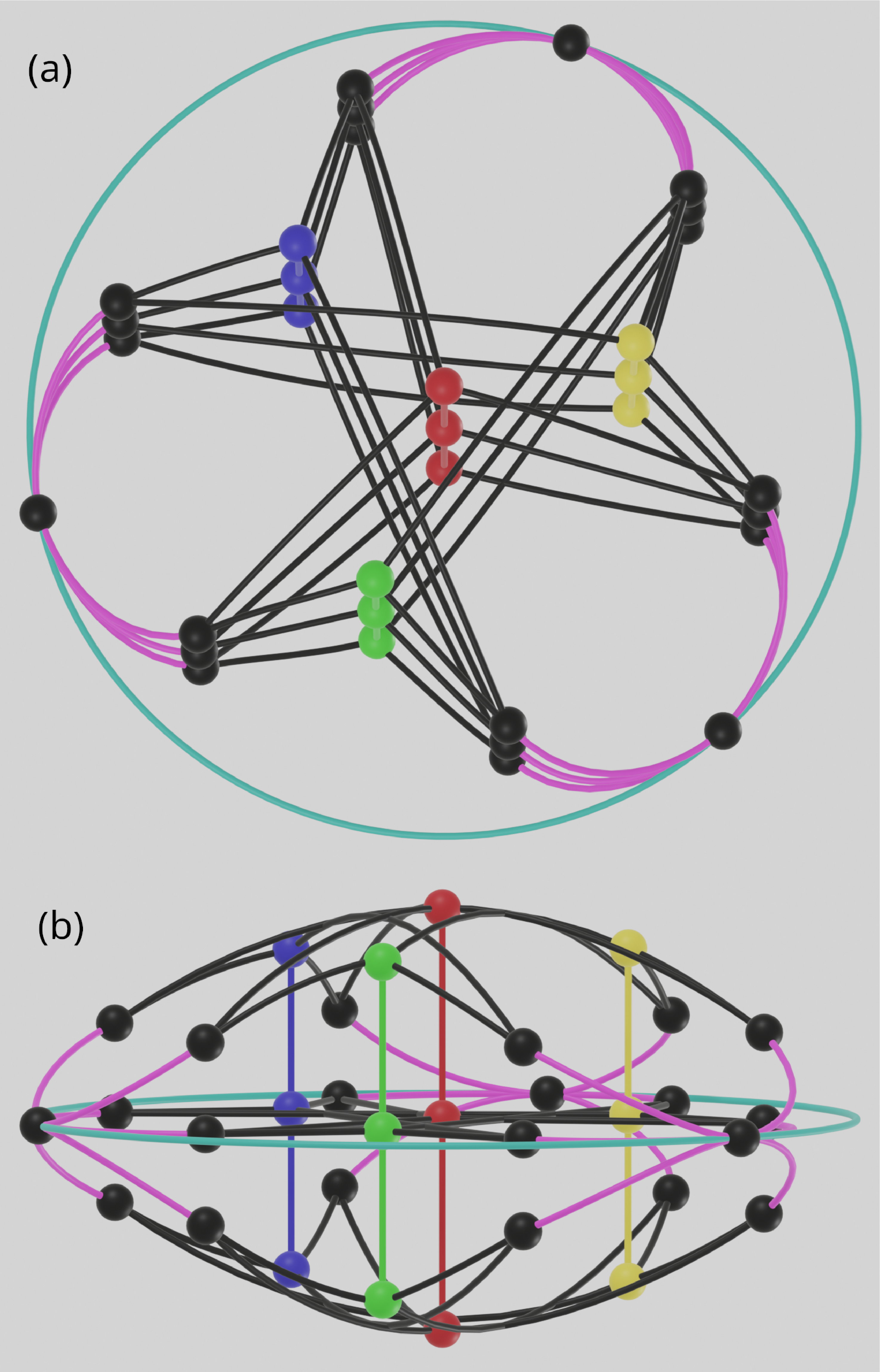}
  \end{center}
  \caption{A 3D representation of the 33-50 set; snapshots from
   two different angles are taken from a Blender output obtained in
   \cite{pavicic-ravlic-2025} which the reader can interactively
   rotate at will; (a) top view; (b) side view.}
\label{fig:yuoh}
\end{figure}

{\em Yu-Oh vs.~KS.\/} \ \  First, there is apparently a claim
\cite[p.~4, top]{cabello-25b} that the Yu-Oh set shown in
Fig.~\ref{fig:CK}(a) is not a KS set (in Cabello's notation).
But, one cannot assign 1 and 0 to its vertices so that the
conditions (i) and (ii) of the aforementioned definition be
satisfied. Hence, it is a KS set in Cabello's own notation
(non-KS in the one of ours). Second, Cabello claims that
``every known small KS set contains the Yu-Oh set.''
Apparently, under ``small KS sets'' he considers Bub
(Sch\"utte), Peres, Conway-Kochen and Kochen-Specker's
sets. As our program SUBGRAPH shows, this is true for the
first three but not for the fourth set. The Yu-Oh 13-16 set
is not a subset of the Kochen-Specker 117-118 set. This shows
that a choice of vector components which generates contextual
sets determines their structure. There is a number of other
master sets (which we obtained in 2022) and their minimal
sets which also do not contain the Yu-Oh set.

{\em Complex vectors.\/} \ \ The 33-50 set in \cite{cabello-25b}
makes use of vector components from $\{0,\pm 1,\pm \omega\}$.
However, there are components which are necessary for a
coordinatization of the whole 69-50 set with the vertices of
multiplicity 1 included. The minimal set of vector components
which generates its master set 157-100 is
$\{0,\pm 1,2\omega,\pm\omega^2,2\omega^2\}$ (notice
$\pm 1$ instead of $\pm\omega$ which gave the master set 169-120).
The coordinatization for the 33-50 can be read off from those of
the 69-50 given in the Appendix (unlike the one of \cite{cabello-25b},
it includes $\omega^2$). It is interesting that the 157-100 contains
only one critical set: the 69-50, while the 169-120 has 514 critical
sets, the biggest of which is 106-79. The 69-50 and the 33-50 have a
high degree of symmetry in a 3D presentation: Fig.~\ref{fig:yuoh}.
In 6D, sets generated by complex vectors have a higher degree of
symmetry than those generated by real
vectors.~\cite[App.~B]{pm-entropy18},
\cite[Fig.~8(c,d)]{pavicic-entropy-25} In 4D, the symmetry is not so
distinct.~\cite[Figs.~5,6]{pavicic-pra-17} In 5D neither. In 7D,
$\{0,\pm\omega\}$ yields a master with 1093 vertices and 9936
hyperedges. A demanding task for a supercomputer.

\begin{figure}[ht]
\begin{center}
  \includegraphics[width=0.49\textwidth]{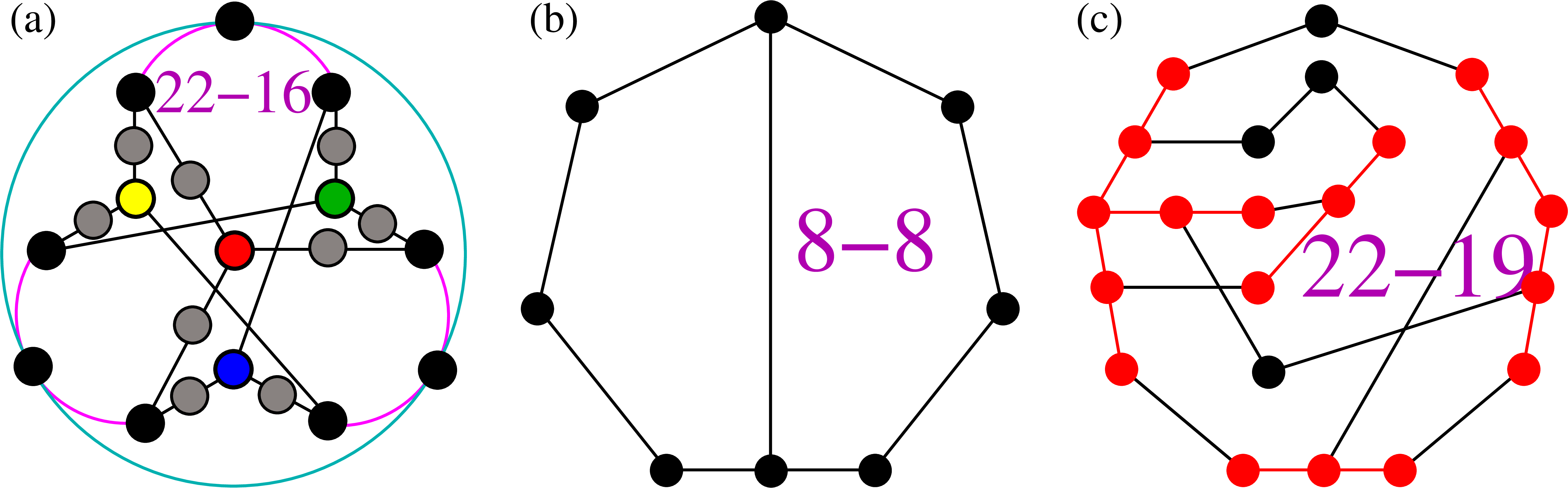}
  \end{center}
  \caption{(a) Partially ``extended'' Yu-Oh set which is still
  contextual and has thirteen complete bases; (b) contextual subset
  of the 33-50 contextual set; (c) contextual sub-hypergraph of
  the 33-50 contextual set with seven complete bases.}
\label{fig:strip}
\end{figure}

{\em Complete bases vs.~KS.\/} \ \ Whether a contextual set has more
or fewer complete bases depends on a choice of vector components used
to generate it and on our decision to drop more or fewer vertices
with multiplicity 1 to achieve or keep the contextuality.
So, for instance, the Yu-Oh set with all but three vertices of
multiplicity 1 added, as shown in Fig.~\ref{fig:strip}(a)
(grey dots), is a KS set (in Cabello's notation, non-KS in
our) with 13 complete bases. Of course, we can limit ourselves
to complete bases without vertices of multiplicity 1 but then
we still have varieties of contextual sub-MMPHs (sub-hypergraphs,
subsets) with different numbers of complete bases as shown in
Fig.~\ref{fig:strip}(b,c) which are KS sets (in Cabello's notation,
non-KS in our). 

\begin{figure}[h]
\begin{center}\vskip10pt
  \includegraphics[width=0.49\textwidth]{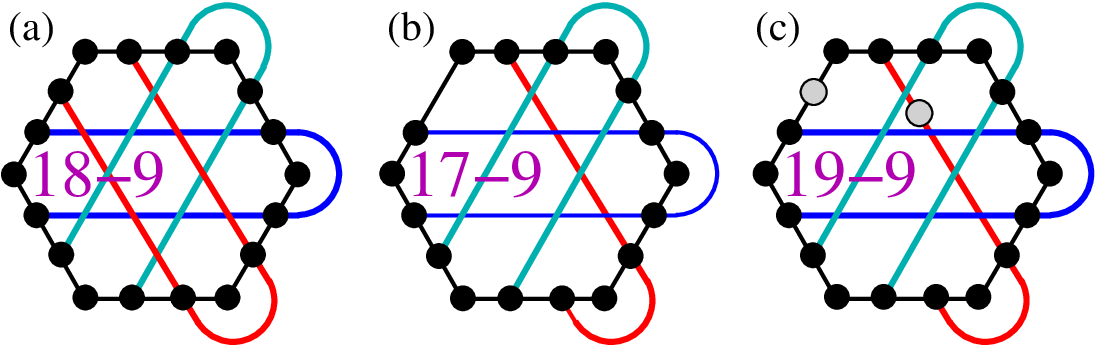}
  \end{center}
  \caption{(a) The 18-9 contextual 4D set
    \cite{cabell-est-96a}, \cite[Fig.~3(a)]{pmmm04c},
    \cite[Fig.~1]{cabello-08};
   (b) the 17-9---a contextual critical subset of 18-9 obtained
   obtained by means of a weak deletion of a vertex
   \cite[Sec.~7.4]{voloshin-09};
   in Cabello's notation it is a KS set while it is
   a non-KS in the notation of ours;
   (c) the 19-9---non-contextual weakly extended set of 17-9.}
\label{fig:ksnonks}
\end{figure}

{\em Notation.} To avoid ambiguities of the Larsson-Cabello notation
one might attempt to define an ``extended'' set as being constructed
by simply adding vertices with multiplicity 1 (weak extension) as
well as by additionally merging vertices with multiplicity 1 (strong
extension) in the original KS sets. That means that one can add
vectors together with their possible orthogonalities or not and that
the obtained set might or might not be contextual in both cases.
Above, we show that in 3D by the example of the Yu-Oh set which is a
KS set in their notation whose partially weakly extended set is
contextual, while completely weakly extended set is not. In 4D, we
show that via two different ``extended'' sets of the critical KS
17-9 set shown in \ref{fig:ksnonks}(b): a weakly extended 17-9---a
non-contextual 19-9 shown in \ref{fig:ksnonks}(c) with two vertices
of multiplicity 1 added and a strongly extended 17-9---a contextual
18-9 set with these two vertices being merged as shown in
Fig.~\ref{fig:ksnonks}(a). Now, the 17-9 is a KS set in Cabello's
notation, but how should we then call the 18-9? An extended KS?
Anyhow, whichever definition we make use of by adding vertices to
a KS set, sometimes we obtain contextual sets and sometimes not. 

{\em Primacy.\/} Hence, there is nothing special in the 33-36
Bob (Sch\"utte), 33-40 Peres, 31-27 Conway-Kochen, 117-118
Kochen-Specker, or 33-50 Pavi{\v c}i{\'c}-Cabello's sets since
they are all non-critical KS sets (in Cabello's notation, non-KS
in our) and therefore contain a number of smaller contextual
sets, i.e., those that do not allow assignment of 1s and 0s to
their vertices in the same way as the original KS sets do.
The sets that are special are 49-36 Bob (Sch\"utte), 57-40 Peres,
51-37 Conway-Kochen, 8 other aforementioned 51-37, 192-118
Kochen-Specker, or 69-50 Pavi{\v c}i{\'c}'s sets since they are
all critical contextual sets---KS sets in our notation---extended
KS sets in Cabello's notation. That is why it is improper to call
the former ones KS sets instead of non-KS sets or whichever other
name. 

{\em Fundamentality. Conclusion.\/} \ \  If we dispensed with
nonlocal games, then the question of the minimal contextual set
with complete bases would have a simple answer: the 7-7 set
\cite[Fig.~6(a)]{pavicic-entropy-25} has just one complete basis
as well as the 8-8 in Fig.~\ref{fig:strip}(b). 
But if we accepted that nonlocal games would have a role in quantum
computation and communication, then we should better consider
how available smaller contextual sets with fewer than 14 complete
bases might be used for the purpose. As we stressed above, there is
no reason to stop at 14 and therefore there is no reason to call
the 33-50 set fundamental. For instance, the 33-50 set contained in
the 69-50 set obtained in \cite{pavicic-quantum-23} and repeated in
\cite{cabello-25b} contains thousands of non-isomorphic
sub-hypergraphs (sub-MMPHs). By removing a chosen number of vertices
with multiplicity 1 from them we obtain a plethora of critical and
non-critical  contextual MMPHs containing fewer than 14 complete
bases which might be considered for nonlocal game designs.

A development of methods of automated generation and analysis of
contextual sets of diverse kinds in the past several decades which
enabled us to unify contextual language, notation, and approaches
and establish a massive database of contextual sets obtained via
supercomputers with the help of our programs leads us to a genesis
of AI contextuality tool which we currently work on. 


Programs are freely available from our
repository \cite{puh-repository}.

\appendix
    \section{Appendix. Coordinatization of the 69-50 in Fig.~\ref{fig:CK}(c)}

{\bf 69-50} {\tt 123,145,267,389,9YA,5ZA,4aB,6bB,7cC,\quad\ \break 8dC,5eD,6fD,8gD,4hC,7iA,9jB,1EF,2GH,3IJ,KkF,\break KlJ,KmH,LnE,LoG,LpJ,MqH,MrI,MsE,NtF,NuG,NvI,\break 1OP,2QR,3ST,UwO,UxT,UyR,VzP,V!Q,V"T,W\#R,W\$S,\break W\%P,X\&O,X'Q,X(S,BLV,CMW,AKU,DNX}.\break {\tt 1}=\{0,0,1\}, {\tt 2}=\{0,1,0\}, {\tt 3}=\{1,0,0\}, {\tt 4}=\{1,-$\omega^2$,0\}, {\tt 5}=\{1,$\omega^2$,0\}, {\tt 6}=\{1,0,-$\omega^2$\}, {\tt 7}=\{1,0,$\omega^2$\}, {\tt 8}=\{0,1,1\},\break {\tt 9}=\{0,1,-1\}, {\tt A}=\{-1,$\omega^2$,$\omega^2$\}, {\tt B}=\{1,$\omega^2$,$\omega^2$\}, {\tt C}=\{1,$\omega^2$,-$\omega^2$\},\break {\tt D}=\{1,-$\omega^2$,$\omega^2$\}, {\tt E}=\{1,-1,0\}, {\tt F}=\{1,1,0\}, {\tt G}=\{$\omega^2$,0,-1\}, {\tt H}=\{$\omega^2$,0,1\}, {\tt I}=\{0,$\omega^2$,1\}, {\tt J}=\{0,$\omega^2$,-1\}, {\tt K}=\{-$\omega^2$,$\omega^2$,1\}, {\tt L}=\{$\omega^2$,$\omega^2$,1\}, {\tt M}=\{$\omega^2$,$\omega^2$,-1\}, {\tt N}=\{$\omega^2$,-$\omega^2$,1\}, {\tt O}=\{$\omega^2$,1,0\}, {\tt P}=\{$\omega^2$,-1,0\}, {\tt Q}=\{1,0,-1\}, {\tt R}=\{1,0,1\}, {\tt S}=\{0,1,$\omega^2$\},\break {\tt T}=\{0,1,-$\omega^2$\}, {\tt U}=\{-$\omega^2$,1,$\omega^2$\}, {\tt V}=\{$\omega^2$,1,$\omega^2$\}, {\tt W}=\{$\omega^2$,1,-$\omega^2$\},\break {\tt X}=\{$\omega^2$,-1,$\omega^2$\}, {\tt Y}=\{2$\omega$,1,1\}, {\tt Z}=\{1,-$\omega^2$,2$\omega^2$\},\break {\tt a}=\{-1,-$\omega^2$,2$\omega^2$\}, {\tt b}=\{-1,2$\omega^2$,-$\omega^2$\}, {\tt c}=\{-1,2$\omega^2$,$\omega^2$\},\break {\tt d}=\{2$\omega$,-1,1\}, {\tt e}=\{-1,$\omega^2$,2$\omega^2$\}, {\tt f}=\{1,2$\omega^2$,$\omega^2$\}, {\tt g}=\{2$\omega$,1,-1\},\break {\tt h}=\{1,$\omega^2$,2$\omega^2$\}, {\tt i}=\{1,2$\omega^2$,-$\omega^2$\}, {\tt j}=\{2$\omega$,-1,-1\}, {\tt k}=\{1,-1,2$\omega$\},\break {\tt l}=\{2$\omega^2$,$\omega^2$,1\}, {\tt m}=\{$\omega^2$,2$\omega^2$,-1\}, {\tt n}=\{-1,-1,2$\omega$\},\break {\tt o}=\{-$\omega^2$,2$\omega^2$,-1\}, {\tt p}=\{2$\omega^2$,-$\omega^2$,-1\}, {\tt q}=\{-$\omega^2$,2$\omega^2$,1\}, {\tt r}=\{2$\omega^2$,-$\omega^2$,1\}, {\tt s}=\{1,1,2$\omega$\}, {\tt t}=\{-1,1,2$\omega$\}, {\tt u}=\{$\omega^2$,2$\omega^2$,1\}, {\tt v}=\{2$\omega^2$,$\omega^2$,-1\}, {\tt w}=\{$\omega^2$,-1,2$\omega^2$\}, {\tt x}=\{2$\omega^2$,1,$\omega^2$\},\break {\tt y}=\{1,2$\omega$,-1\}, {\tt z}=\{-$\omega^2$,-1,2$\omega^2$\}, {\tt !}=\{-1,2$\omega$,-1\},\break {\tt "}=\{2$\omega^2$,-1,-$\omega^2$\}, {\tt \#}=\{-1,2$\omega$,1\}, {\tt \$}=\{2$\omega^2$,-1,$\omega^2$\}, {\tt \%}=\{$\omega^2$,1,2$\omega^2$\}, {\tt \&}=\{-$\omega^2$,1,2$\omega^2$\}, {\tt '}=\{1,2$\omega$,1\}, {\tt (}=\{2$\omega^2$,1,-$\omega^2$\}

%


\end{document}